# Raman scattering and anomalous Stokes–anti-Stokes ratio in MoTe$_2$ atomic layers

*Thomas Goldstein[1†], Shao-Yu Chen[1†], Di Xiao[2], Ashwin Ramasubramaniam[3] and Jun Yan[1*]*

[1]Department of Physics, University of Massachusetts, Amherst, Massachusetts 01003, USA

[2]Department of Physics, Carnegie Mellon University, Pittsburgh, PA 15213, USA

[3]Department of Mechanical & Industrial Engineering, University of Massachusetts, Amherst, Massachusetts 01003, USA

[†]These authors contributed equally to this work.

[*]Corresponding Author: Jun Yan.    Tel:    (413)545-0853    Fax:    (413)545-1691
E-mail: yan@physics.umass.edu

**Abstract**

Stokes and anti-Stokes Raman scattering are performed on atomic layers of hexagonal molybdenum ditelluride (MoTe$_2$), a prototypical transition metal dichalcogenide (TMDC) semiconductor. The data reveal all six types of zone center optical phonons, along with their corresponding Davydov splittings, which have been challenging to see in other TMDCs. We discover that the anti-Stokes Raman intensity of the low energy layer-breathing mode becomes more intense than the Stokes peak under certain experimental conditions, and find the effect to be tunable by excitation frequency and number of atomic layers. These observations are interpreted as a result of resonance effects arising from the C excitons in the vicinity of the Brillouin zone center in the photon-electron-phonon interaction process.

**Introduction**

The coupling between photons, electrons, and phonons is central to understanding fundamental properties of condensed matter systems. In layered transition metal dichalcogenide (TMDC) semiconductors, electron-photon coupling enables investigation of the under-screened strong Coulomb interaction in reduced dimensions[1] as well as the manipulation of spins and pseudo-spins[2,3]. The interaction of chiral phonons with electrons and circularly polarized photons is linked to pseudo-spin flip, and may lead to a novel valley phonon Hall effect, as predicted by a recent theory[4]. Despite much recent experimental progress[2,5], the coupled photon-electron-phonon system in TMDCs is still not fully understood, and continues to be a focus of two dimensional (2D) materials research.

In this work we study such interaction effects in molybdenum ditelluride ($MoTe_2$), a prototypical TMDC semiconductor, which has recently attracted a great deal of interest. Investigations of multi- and mono-layer $MoTe_2$ transistor devices[6–8] have probed the semiconductor's band gap and transport properties, revealing ambipolar charge conduction channels, and a stable metallic T' phase which is potentially a type-II Weyl semimetal[9,10], and can be used to make Ohmic homojunction contacts[6]. Optically, monolayer $MoTe_2$ features strong photoluminescence similar to the other TMDCs, but it has by far the smallest optical bandgap at the K points, lying in the infrared around 1 eV[11,12]. As a result, the C excitons located near the Brillouin zone center ($\Gamma$)[13] are in the visible range around 2.5eV[11], which we take advantage of to investigate the resonance effects reported here.

We use Raman Spectroscopy to probe the photon-electron-phonon coupling in $MoTe_2$, revealing hard to access Raman modes and mode splitting in detail. We also observe a peculiar phenomenon in which the anti-Stokes Raman peak becomes more intense than the Stokes peak, which is contrary to the usual intuition from Boson statistics, and could potentially serve as a laser cooling channel for the atomically thin 2D crystals. Our further investigation into how the photon-electron-phonon coupling affects the anti-Stokes intensity reveals that the phenomenon is tunable according to laser wavelength and MoTe$_2$ layer number. These experimental observations highlight the role of the C excitons, and provide insights into the lattice dynamics and electronic structures of molybdenum ditelluride.

**Results and Discussion**

The MoTe$_2$ atomic layers used in this work are exfoliated from bulk crystals grown via chemical vapor transport with chlorine as the transport agent, as illustrated in Figure 1a (more details in Methods). Typical sizes of the crystals (Figure 1b) range from a few mm to 1 cm. Atomic layers of MoTe$_2$ are exfoliated with Scotch tape and deposited on silicon with a 300 nm oxide layer or on fused silica substrates. Figure 1c shows the optical microscope image of a typical atomic flake with one to five layers of MoTe$_2$. Atomic force microscope (AFM) characterization in Figure 1d shows that each additional three-atom-thick MoTe$_2$ layer has a step height of about 0.7nm, consistent with previous measurements[11].

Atomic layers of hexagonal TMDCs host six prototypical zone-center optical phonons, shown in the bilayer dispersion (see Supporting Information for details of DFT calculation) and atomic displacement drawings of Figure 2a and 2b, respectively. These include the shear mode (S), the breathing mode (B), the in-plane chalcogen vibration (IC), the out-of-plane chalcogen vibration (OC), the in-plane metal-chalcogen vibration (IMC), and the out-of-plane metal-chalcogen vibration (OMC)[14]. We measure the in-plane S, IC and IMC modes with cross polarization $\bar{z}(xy)z$, and the out-of-plane B, OC and OMC modes in parallel polarization $\bar{z}(xx)z$ (see Methods for details of Raman set-up). Figure 2c shows the Raman spectra excited with the 532 nm (2.33 eV) laser light. The mode energies, in cm$^{-1}$ (1meV ≈ 8.06 cm$^{-1}$), are given in Table 1; for the bilayer, values derived from DFT are given in parenthesis for comparison.

The occurrence and splitting of the observed Raman peaks can be compared with group theoretical predictions: Table 2 lists the symmetries and expected number of modes for the six types of zone-center optical phonons in one to five layers and in the bulk as derived from Ref. [14]. Except for the bulk B and OMC modes of $B_{2g}$ symmetry[15], all modes that are even under inversion/mirror-reflection operation ($i/\sigma_h$) are Raman active (bold red text in Table 2). For 1L-MoTe$_2$ the S and B modes are absent since interlayer vibrations only occur in multi-layers, and the IC and OMC modes are not Raman active; this is in agreement with the observation of only the OC and IMC modes in the monolayer (black) spectra in Figure 2c. Bilayer MoTe$_2$ has one Raman active mode for each of the six types of zone center optical phonons, consistent with the six peaks observed in bilayer (red) spectra.

For samples thicker than bilayer the Raman spectra display Davydov splitting due to interlayer interactions. We observe two B and two S modes in 4L and 5L samples; two OC modes in 3L and 4L samples; and three OC modes in 5L samples, in perfect agreement with the theoretical expectations in Table 2. The IC mode appears to display non-monotonic dependence on layer numbers, and develops an asymmetric shape for 4L and 5L samples. This indicates that the IC spectra in 4L and 5L samples are composed of two peaks as suggested by the theory in Table 2. With this interpretation (dashed fitting curves of Figure 2c), the IC vibrations are composed of two sets of modes consistently red-shifting with increasing number of layers. The IMC is expected to split in a manner similar to OC, and OMC similar to IC, but neither were observed experimentally: IMC and OMC display a single symmetric peak in multi-layer MoTe$_2$ in Figure 2c. This might be due to either the splitting being small compared with the linewidth or the split modes having too small Raman cross-section. We note that the former interpretation is in agreement with the theoretical phonon dispersion calculation in Figure 2a where the even ($A_{1g}$) and odd ($A_{2u}$) OC vibrations in 2L-MoTe$_2$ have the largest Davydov splitting.

In the other four hexagonal TMDCs, (Mo,W)(S,Se)$_2$, the OC and IMC modes are the dominant Raman features, and the B and S modes become observable if low-energy stray light is cut, but the IC and OMC modes tend to evade most Raman scattering measurements[14,16–22] unless less-common ultraviolet lasers are used[17]. However, because of MoTe$_2$'s small optical band gap[11], the visible lasers used here can access the C excitons arising from electron states in the vicinity of the Γ point[23]. Recent theoretical calculations[24] show that the C exciton is composed of six nearly degenerate states involving the topmost valence band and the three lowest conduction bands, in contrast to the A and B excitons where only one conduction and one valence band near the K points are involved. The availability of multiple bands and states provides more electronic degrees of freedom for electron-phonon interaction, which in turn assist the excitation of all of the zone-center optical phonons in MoTe$_2$. In capturing all six zone center modes, our Raman spectra in Figure 2c and Table 1 constitute a comprehensive data set for TMDC lattice dynamics, in agreement with another recent work on MoTe$_2$ [25].

In addition to enabling observation of all six branches of zone center lattice vibrations, the C excitons in the photon-electron-phonon interaction Raman process have interesting impacts on Raman intensity. In particular we have found that the resonances have dramatic effects on the Stokes and anti-Stokes Raman signatures of the low energy breathing modes. Figure 3a shows the breathing mode Stokes and anti-Stokes peaks for 2 through 5 layers, using the 2.41 eV laser excitation. (The Raman intensities in Figures 3 and 4 are calibrated; see Supporting Information.) As expected, the anti-Stokes peaks are distributed symmetrically in Raman shift about zero with respect to the Stokes peaks. However we also observe that the anti-Stokes peaks are more intense than the Stokes peaks, an unusual occurrence not present in most systems.

In a typical, lowest-order Raman scattering process for zone-center optical phonons, there are three steps involved as shown in Figure 3b:

1. The electronic system absorbs a photon and creates an electron-hole pair;

2. The electron-hole pair creates (Stokes) or absorbs (anti-Stokes) a phonon;

3. The electron-hole pair recombines and emits the scattered photon.

The first and the third steps are linked to electron-photon interaction while the second step is determined by electron-phonon interaction. The rate of step 2 is proportional to the phonon creation or destruction operator acting on the system squared, giving a factor of ($n + 1$) or $n$, respectively, where $n$ is the phonon occupation number. As phonons are bosonic excitations with zero chemical potential, $n = \frac{1}{e^{\frac{\hbar\omega}{k_BT}}-1}$, where $\hbar\omega$ is the phonon energy and $k_BT$ is the thermal energy. For step 3, the recombining electron-hole pair acts similarly to a radiating dipole, whose emission rate carries a factor of $\omega^4$ according to classical field theory[26]. Thus, the anti-Stokes to Stokes intensity ratio is predicted to be

$$\frac{I_{as}}{I_s} = \frac{\omega_{as}^4}{\omega_s^4}\frac{n}{n+1} = \frac{\omega_{as}^4}{\omega_s^4}e^{-\frac{\hbar\omega}{k_BT}} \qquad (1).$$

Note that the impact of the polynomial $\frac{\omega_{as}^4}{\omega_s^4}$ is typically much smaller than the exponential $e^{-\frac{\hbar\omega}{k_BT}}$; thus $I_{as}$ is expected to be less intense than $I_s$, a condition which is ubiquitously observed in most material systems.

The anomalously intense anti-Stokes peaks suggest that some other factor not included in Equation (1) plays an important role in determining the Raman cross section. Step 1 in Figure 3b is shared by the Stokes and anti-Stokes process, and the differences of step 2 are purely in occupation number, so neither can explain the anti-Stokes intensity. This leaves elements of step 3 not included in the classical theory of radiating dipoles; in particular any resonances between the final emitted Stokes and anti-Stokes photons and the electronic structure of the crystal. For our frequency ranges this is, as mentioned above, the C exciton, which will have a different resonance effect on the Stokes and anti-Stokes emitted photons due to their different energies.

The resonance effect becomes even clearer when we systematically measure the calibrated breathing mode intensities using different laser lines. Figure 4a shows Raman data (normalized by Stokes intensities) for 5L-MoTe$_2$ excited by 488 nm (2.54 eV), 514 nm (2.41 eV), and 532 nm (2.33 eV) respectively. The anti-Stokes to Stokes intensity ratio depends sensitively on the incident laser excitation: at 2.33 eV, the anti-Stokes peak has about the same intensity as the Stokes peak; at 2.41 eV and 2.54 eV, the anti-Stokes intensity is above and below the Stokes, respectively. From these observations, we introduce a resonance correction factor, $R$, into Equation (1), such that

$$\frac{I_{as}}{I_s} = R \frac{\omega_{as}^4}{\omega_s^4} \frac{n}{n+1} = R \frac{\omega_{as}^4}{\omega_s^4} e^{-\frac{\hbar\omega}{k_B T}} \qquad (2).$$

In Figure 4b we plot $ln\left(\frac{\omega_s^4}{\omega_{as}^4} \frac{I_{as}}{I_s}\right) = lnR - \frac{\hbar\omega}{k_B T}$ as a function of the breathing mode phonon energy for 2 through 5 layer MoTe$_2$ with all three excitation wavelengths. In the absence of any correction, i.e. $R=1$ in Equation (2), the experimental data are expected to fall within the gray band, where the two bounding dotted black lines specify the temperature uncertainty during our experiment (see Supporting Information). The experimental data show significant deviation from $R=1$, with the enhancement and suppression of the anti-Stokes to Stokes ratio mostly consistent across layer number.

To further establish the connection between the C exciton and the laser-wavelength / atomic-layer-number dependent $R$ factor, we collected differential reflection data, presented in Figure 4c, showing the C exciton resonance for 2 through 5 layer MoTe$_2$ (see Methods). The data strongly imply that the observed deviations of the $R$ factor from unity are correlated with the overall shape and evolution of the C exciton: the measured $ln\left(\frac{\omega_s^4}{\omega_{as}^4} \frac{I_{as}}{I_s}\right)$ for 5L samples has the largest variations at different laser excitations, consistent with the fact that its exciton resonance is the most sharply peaked. For 3L to 5L samples, 2.54 eV lies on the opposite side of the exciton peak from 2.41 and 2.33 eV; this corresponds to the swapping from resonance suppression to resonance enhancement of the anti-Stokes intensity. The slopes of the differential reflection for 2.33 and 2.41 eV are roughly constant; this corresponds to the roughly constant $R$ value across different numbers of layers for these energies. In contrast the slopes for 2.54 eV excitations have significant variations, as do the resonance effects in different numbers of layers.

We finally remark that while other materials with exciton effects may also exhibit an anomalous Stokes–anti-Stoke ratio, it is quite rare to have the anti-Stokes peak actually be more intense than the Stokes peak. Also we take note that the Stokes–anti-Stokes photon energy difference for the 5L-MoTe$_2$ breathing mode is only about 3 meV, corresponding to about 0.15% of the total photon energy. It is quite remarkable that the C exciton resonance at room temperature can amplify this minute change in energy to switch the relative Stokes–anti-Stokes Raman cross-section. The reversed intensity ratio in fact indicates that more phonons are removed from, rather than created in, the crystal, potentially forming a laser-cooling channel in the system. Indeed, the observation of intense anti-Stokes in another semiconductor system (CdS nanoribbons) is accompanied with strong laser cooling effects[27]. We thus surmise that our observations point to the uncharted field of laser cooling of atomically thin 2D crystals, corroborating a recent study of luminescence upconversion in WSe$_2$[28].

**Conclusion**

We have exfoliated few layer MoTe$_2$ from chemical vapor transport grown crystals. Using polarization-resolved Raman scattering we have unambiguously identified all six zone-center optical phonons. Splitting of these phonons, where observed, has the correct theoretically predicted number of branches. We discovered that the anti-Stokes intensities deviate from their expected values, to the point of being more intense than the Stokes intensities, with the intensity ratio tunable via resonance with the C exciton. A stronger anti-Stokes peak implies that more phonons are annihilated than created in the crystal suggesting that, from the electron-photon-phonon interaction process, anti-Stokes Raman scattering could provide a cooling channel in atomically thin TMDC crystals.

## Methods

**Bulk MoTe$_2$ crystal growth**

We grew bulk MoTe$_2$ crystals using chemical vapor transport with chlorine as the transport agent, as illustrated in Figure 1a. A horizontal three-zone Carbolite furnace provides a high temperature reaction zone and a low temperature growth zone. Following the work of Ubaldini et al.[29], Mo, Te, and MoCl$_5$ powders are placed in a fused silica tube, 18 mm in diameter and 300 mm in length. The purity of the source materials are Mo 99.9%, Te 99.997%, and MoCl$_5$ 95% (Sigma Aldrich). Total Mo and Te are kept in a stoichiometric 2:1 ratio with sufficient MoCl$_5$ to achieve a Cl density of 2 mg/cm$^3$. The tube is pump-purged with ultra-high purity argon gas and sealed at low pressure prior to growth. The reaction and growth zones were kept at 830ºC and 730ºC respectively for 100 hours. During cooling the temperature profile is inverted with the growth zone being roughly 100ºC hotter than the reaction zone so that chloride contaminants preferentially precipitate in the hot zone, away from the MoTe$_2$ crystals.

**Raman Scattering and Differential Reflection**

We collect Raman data using linearly-polarized light from an Argon laser or a frequency doubled Nd:YAG solid state laser to excite the samples in a back scattering geometry. The scattered light is collected with a 100x objective lens, dispersed by a triple stage spectrometer, and detected by a liquid nitrogen cooled CCD. In the collection path, a broadband polarization rotator and a linear polarizer are used to selectively detect scattered light with polarization either parallel $\bar{z}(xx)z$ or perpendicular $\bar{z}(xy)z$ to that of the incident light. With a combination of a Bragg diffraction grating (OptiGrate) and a Horiba T64000 triple stage spectrometer operating in the subtractive mode, we were able to observe low energy phonon modes down to less than 10 cm$^{-1}$.

We obtain differential reflection data by exfoliating MoTe$_2$ atomic layers onto fused silica. We then shine a white laser (NKT Photonics) on the samples and collect the reflection spectrum, which is subtracted from and normalized by the substrate reflection.

**Author contributions:**

T. G. and S.-Y. C. contributed equally to this work. J.Y. conceived the experiment. T. G. grew the MoTe$_2$ crystals. T.G. and S.-Y. C. exfoliated the atomic layers, performed the measurements and carried out the data analysis. A.R. calculated the bilayer MoTe$_2$ phonon dispersion. All authors contributed to writing the manuscript.

**Acknowledgement**:

This work is supported by the University of Massachusetts Amherst, the National Science Foundation Center for Hierarchical Manufacturing (CMMI-1025020) and Office of Emerging Frontiers in Research and Innovation (EFRI-1433496). Computing support from the Massachusetts Green High Performance Computing Center is gratefully acknowledged.

**Competing financial interests:** The authors declare no competing financial interests.

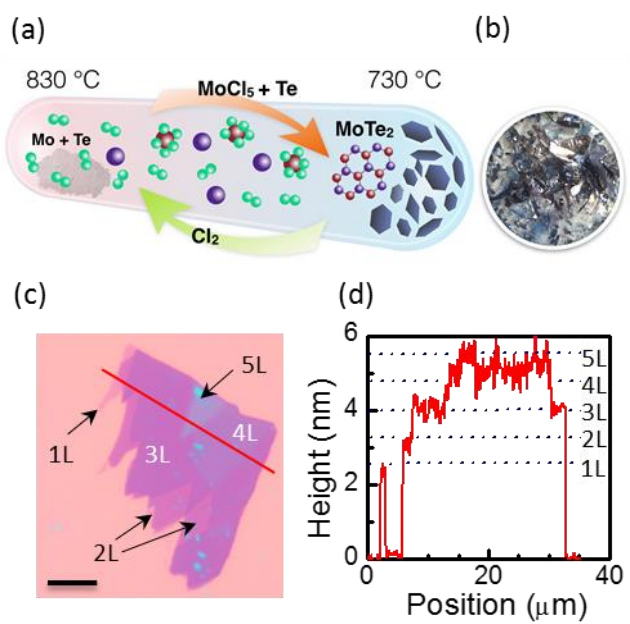

**Figure 1.** (a) Schematic illustration of the chemical vapor transport crystal growth process. (b) Image of grown $MoTe_2$ crystals. (c) Optical microscope image and (d) AFM trace [along the red line in (c)] of a mechanically exfoliated $MoTe_2$ flake consisting of 1 to 5 atomic layers.

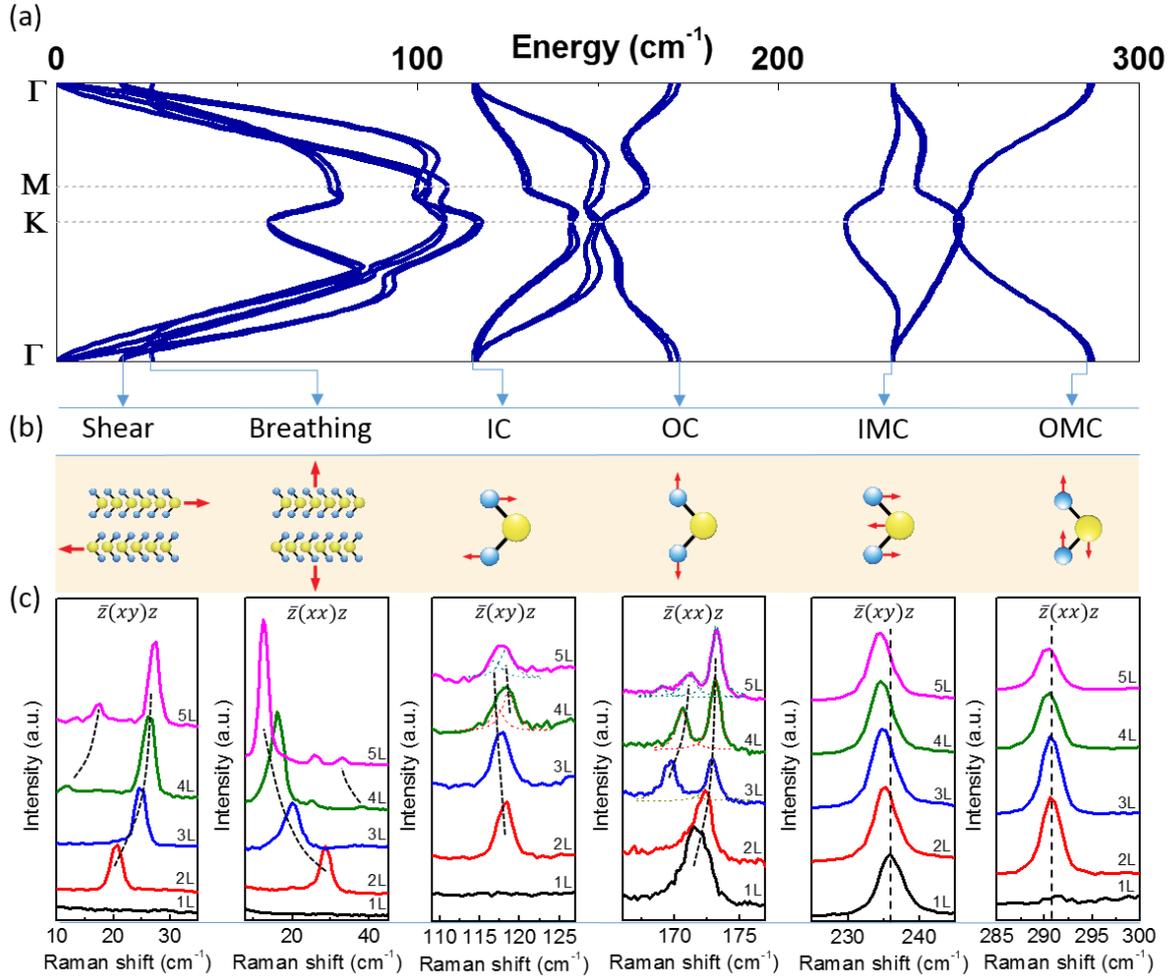

**Figure 2.** (a) Phonon dispersion of bilayer MoTe$_2$ calculated with density functional theory (DFT). (b) Atomic displacements of the six prototypical TMDC Γ point optical phonons. The arrows between (a) and (b) connect the calculated Γ point optical phonons with the names we use to identify the lattice vibrations. (c) Experimental Raman spectra for all six phonon branches in 1L to 5L MoTe$_2$; the spectra have been vertically shifted for clarity. The black dashed curves are guides to the eye.

| Mode | 1L | 2L | 3L | 4L | 5L |
|---|---|---|---|---|---|
| S | | 19.6 (17.9) | 24 | 10.4 / 25.3 | 16.7 / 26.2 |
| B | | 28.6 (26.5) | 20.9 | 16.2 / 37.9 | 12.6 / 32.9 |
| IC | | 118.1 (116.5) | 117.6 | 118.1 | 117.8 |
| OC | 171.5 | 172.4 (172.3) | 172.9 / 169.6 | 173.1 / 170.5 | 173.2 / 171.1 / 169.0 |
| IMC | 236.0 | 235.2 (231.7) | 234.9 | 234.7 | 234.5 |
| OMC | | 290.7 (287.2) | 290.6 | 290.4 | 290.3 |

**Table 1.** Energy (cm$^{-1}$) of the observed zone-center optical phonon modes from 1L to 5L MoTe$_2$. For the bilayer, DFT results are included in the parentheses.

| # of Layer | Sym. Grp. | $\sigma_h$/i Sym. | S | B | IC | OC | IMC | OMC |
|---|---|---|---|---|---|---|---|---|
| 1 | $D_{3h}$ | + | - | - | - | 1 A$'_1$ | 1 E$'$ | - |
| | | − | - | - | 1 E$''$ | - | - | 1 A$''_2$ |
| 2 | $D_{3d}$ | + | **1 E$_g$** | **1 A$_{1g}$** | **1 E$_g$** | **1 A$_{1g}$** | **1 E$_g$** | **1 A$_{1g}$** |
| | | − | - | - | 1 E$_u$ | 1 A$_{2u}$ | 1 E$_u$ | 1 A$_{2u}$ |
| 3 | $D_{3h}$ | + | **1 E$'$** | **1 A$'_1$** | **1 E$'$** | **2 A$'_1$** | **2 E$'$** | **1 A$'_1$** |
| | | − | 1 E$''$ | 1 A$''_2$ | 2 E$''$ | 1 A$''_2$ | 1 E$''$ | 2 A$''_2$ |
| 4 | $D_{3d}$ | + | **2 E$_g$** | **2 A$_{1g}$** | **2 E$_g$** | **2 A$_{1g}$** | **2 E$_g$** | **2 A$_{1g}$** |
| | | − | 1 E$_u$ | 1 A$_{2u}$ | 2 E$_u$ | 2 A$_{2u}$ | 1 E$_u$ | 2 A$_{2u}$ |
| 5 | $D_{3h}$ | + | **2 E$'$** | **2 A$'_1$** | **2 E$'$** | **3 A$'_1$** | **3 E$'$** | **2 A$'_1$** |
| | | − | 2 E$''$ | 2 A$''_2$ | 3 E$''$ | 2 A$''_2$ | 2 E$''$ | 3 A$''_2$ |
| bulk | $D^4_{6h}$ | + | **1 E$_{2g}$** | 1 B$_{2g}$ | **1 E$_{1g}$** | **1 A$_{1g}$** | **1 E$_{2g}$** | 1 B$_{2g}$ |
| | | − | - | - | 1 E$_{2u}$ | 1 B$_{1u}$ | 1 E$_{1u}$ | 1 A$_{2u}$ |

**Table 2.** Symmetry representation for phonon modes in bulk and few layer TMDCs[14]. The third column specifies whether the mode is even or odd under horizontal mirror plane reflection/inversion. Raman active modes (in back scattering geometry) are colored in bold red.

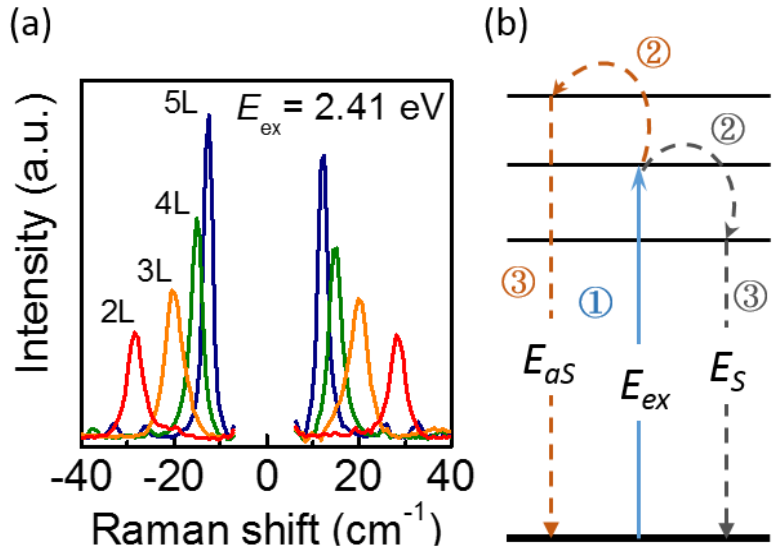

**Figure 3.** (a) Experimental Raman data of MoTe$_2$ under a 2.41 eV excitation, showing the evolution of the breathing mode peak position and intensity with layer number. The anti-Stokes intensity is anomalously greater than the Stokes intensity. (b) Illustration of the zone-center optical-phonon Raman scattering process.

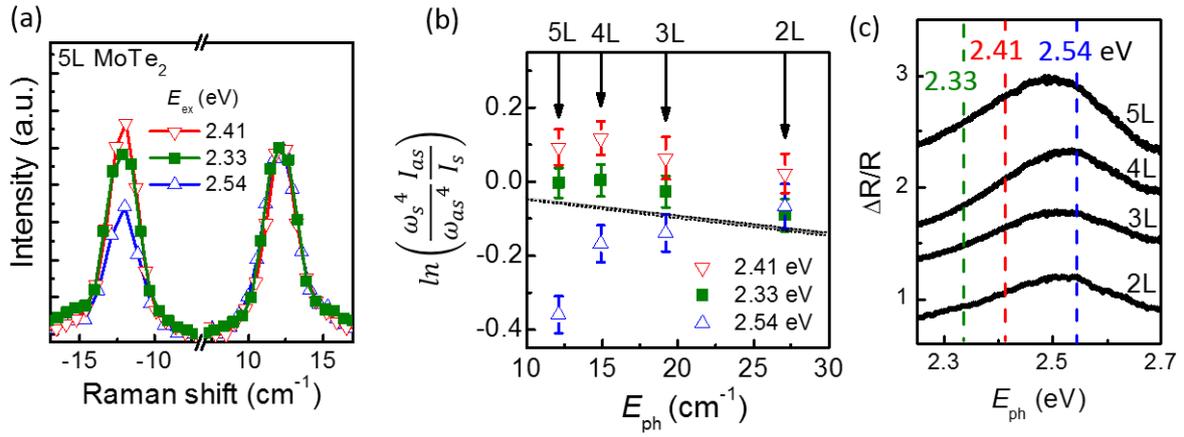

**Figure 4.** (a) Experimental Raman data for 5 layer MoTe$_2$ using 2.33, 2.41, and 2.54 eV excitations. Stokes intensities have been normalized, showing the difference in anti-Stokes intensity between different excitation energies. (b) $ln\left(\frac{\omega_s^4}{\omega_{as}^4}\frac{I_{as}}{I_s}\right)$ as a function phonon mode energy in cm$^{-1}$. Each energy corresponds to a different layer number: 5 layer to 2 layer, from low to high energy. The grey band shows the expected value if there were no resonance effects, with the bounding black dotted lines determined by the experimental temperature uncertainty in our measurements (see Supporting Information). (c) Differential reflectance data for 2 layer to 5 layer MoTe$_2$ in the energy region of the C exciton. Dashed lines indicate the energy of the laser excitations used for Raman measurements in panels (a) and (b).

**References**:

1. Mak, K. F. *et al.* Tightly bound trions in monolayer MoS2. *Nat. Mater.* **12,** 207–211 (2013).

2. Xu, X., Yao, W., Xiao, D. & Heinz, T. F. Spin and pseudospins in layered transition metal dichalcogenides. *Nat. Phys.* **10,** 343–350 (2014).

3. Xiao, D., Liu, G. Bin, Feng, W., Xu, X. & Yao, W. Coupled spin and valley physics in monolayers of MoS2 and other group-VI dichalcogenides. *Phys. Rev. Lett.* **108,** 196802 (2012).

4. Zhang, L. & Niu, Q. Chiral Phonons at High-Symmetry Points in Monolayer Hexagonal Lattices. *Phys. Rev. Lett.* **115,** 115502 (2015).

5. Jariwala, D., Sangwan, V. K., Lauhon, L. J., Marks, T. J. & Hersam, M. C. Emerging device applications for semiconducting two-dimensional transition metal dichalcogenides. *ACS Nano* **8,** 1102–1120 (2014).

6. Cho, S. *et al.* Phase patterning for ohmic homojunction contact in MoTe2. *Science* **349,** 625–628 (2015).

7. Lin, Y. F. *et al.* Ambipolar MoTe2 transistors and their applications in logic circuits. *Adv. Mater.* **26,** 3263–3269 (2014).

8. Pradhan, N. R. *et al.* Field-effect transistors based on few-layered α-MoTe2. *ACS Nano* **8,** 5911–5920 (2014).

9. Sun, Y., Wu, S., Ali, M. N., Felser, C. & Yan, B. Prediction of Weyl semimetal in orthorhombic MoTe 2. *Phys. Rev. B* **92,** 161107(R) (2015).

10. Chen, S.-Y., Goldstein, T., Ramasubramaniam, A. & Yan, J. Inversion-symmetry-breaking-activated shear Raman bands in T'-MoTe2. (2016). at <http://arxiv.org/abs/1602.03566>

11. Ruppert, C., Aslan, O. B. & Heinz, T. F. Optical properties and band gap of single- and few-layer MoTe2 crystals. *Nano Lett.* **14,** 6231–6236 (2014).

12. Lezama, I. G. *et al.* Indirect-to-Direct Band Gap Crossover in Few-Layer MoTe2. *Nano Lett.* **15,** 2336–2342 (2015).

13. Qiu, D. Y., da Jornada, F. H. & Louie, S. G. Optical spectrum of MoS2: Many-body effects and diversity of exciton states. *Phys. Rev. Lett.* **111,** 216805 (2013).

14. Chen, S.-Y., Zheng, C., Fuhrer, M. S. & Yan, J. Helicity-resolved Raman scattering of MoS2, MoSe2, WS2, and WSe2 atomic layers. *Nano Lett.* **15,** 2526–2532 (2015).

15. Yamamoto, M. *et al.* Strong enhancement of Raman scattering from a bulk-inactive vibrational mode in few-layer MoTe2. *ACS Nano* **8,** 3895–3903 (2014).

16. Lee, C. *et al.* Anomalous lattice vibrations of single- and few-layer MoS2. *ACS Nano* **4,** 2695–2700 (2010).

17. Li, H. *et al.* From Bulk to Monolayer MoS2: Evolution of Raman Scattering. *Adv. Funct. Mater.* **22,** 1385–1390 (2012).

18. Berkdemir, A. *et al.* Identification of individual and few layers of WS2 using Raman spectroscopy. *Sci. Rep.* **3,** 1755 (2013).

19. Zhao, Y. *et al.* Interlayer breathing and shear modes in few-trilayer MoS2 and WSe2. *Nano Lett.* **13,** 1007–1015 (2013).

20. Lui, C. H., Ye, Z., Keiser, C., Barros, E. B. & He, R. Stacking-dependent shear modes in trilayer graphene. *Appl. Phys. Lett.* **106,** 041904 (2015).

21. Scheuschner, N., Gillen, R., Staiger, M. & Maultzsch, J. Interlayer resonant Raman modes in few-layer MoS2. *Phys. Rev. B* **91,** 235409 (2015).

22. Lee, J.-U., Park, J., Son, Y.-W. & Cheong, H. Anomalous excitonic resonance Raman effects in few-layered MoS2. *Nanoscale* **7,** 3229–3236 (2015).

23. Bromley, R. A., Murray, R. B. & Yoffe, A. D. The band structures of some transition metal dichalcogenides. III. Group VIA: trigonal prism materials. *J. Phys. C Solid State Phys.* **5,** 759–778 (2001).

24. Qiu, D. Y., da Jornada, F. H. & Louie, S. G. Optical Spectrum of MoS2: Many-Body Effects and Diversity of Exciton States. *Phys. Rev. Lett.* **111,** 216805 (2013).

25. Froehlicher, G. *et al.* Unified Description of the Optical Phonon Modes in N-Layer MoTe2. *Nano Lett.* **15,** 6481–6489 (2015).

26. Cardona, M. *Light Scattering in Solids II*. *Springer-Verlag Berlin Heidelberg* (1982).

doi:10.1080/713820804

27. Zhang, J., Li, D., Chen, R. & Xiong, Q. Laser cooling of a semiconductor by 40 kelvin. *Nature* **493,** 504–508 (2013).

28. Jones, A. M. *et al.* Excitonic luminescence upconversion in a two-dimensional semiconductor. *Nat. Phys.* (2015). doi:10.1038/nphys3604

29. Ubaldini, A. & Jacimovic, J. Chloride-Driven Chemical Vapor Transport Method for Crystal Growth of Transition Metal Dichalcogenides. *Cryst. Growth Des.* **13,** 4453–4459 (2013).